# Compression of structured high-throughput sequencing data

Fabien Campagne[1,2]*, Kevin C. Dorff[1], Nyasha Chambwe[1,2], James T. Robinson[3], Jill P. Mesirov[3], Thomas D. Wu[4].

[1]The HRH Prince Alwaleed Bin Talal Bin Abdulaziz Alsaud Institute for Computational Biomedicine. [2]Department of Physiology and Biophysics, The Weill Cornell Medical College, New York, NY, USA. [3]Broad Institute of Massachusetts Institute of Technology and Harvard, Cambridge, MA, USA. [4]Department of Bioinformatics, Genentech, Inc., 1 DNA Way, South San Francisco, CA, USA.

*Correspondence to: Fabien Campagne, fac2003@campagnelab.org.

**Abstract**: Large biological datasets are being produced at a rapid pace and create substantial storage challenges, particularly in the domain of high-throughput sequencing (HTS). Most approaches currently used to store HTS data are either unable to quickly adapt to the requirements of new sequencing or analysis methods (because they do not support schema evolution), or fail to provide state of the art compression of the datasets. We have devised new approaches to store HTS data that support seamless data schema evolution and compress datasets substantially better than existing approaches. Building on these new approaches, we discuss and demonstrate how a multi-tier data organization can dramatically reduce the storage, computational and network burden of collecting, analyzing, and archiving large sequencing datasets. For instance, we show that spliced RNA-Seq alignments can be stored in less than 4% the size of a BAM file with perfect data fidelity. Compared to the previous compression state of the art, these methods reduce dataset size more than 20% when storing gene expression and epigenetic datasets. The approaches have been integrated in a comprehensive suite of software tools (http://goby.campagnelab.org) that support common analyses for a range of high-throughput sequencing assays.

Many scientific disciplines, including high-energy physics, astronomy and more recently biology, generate increasing volumes of data from automated measurement instruments. In biology, modern high-throughput sequencers (HTS) are producing a large fraction of new biological data and are being successfully applied to study genomes, transcriptomes, epigenomes or other data modalities with a variety of new assays that take advantage of the throughput of sequencing methods[1-3]. In addition to sequenced reads, data analyses yield secondary data, such as alignments of reads to reference genome. Sequencing throughput has more than doubled every year for the last ten years[4] resulting in storage requirements on the order of tens of terabytes of primary and secondary high throughput sequencing data in a typical laboratory. Major sequencing centers in the USA and worldwide typically require several tens of petabytes of storage to store reads and secondary data during the lifetime of their projects. Before study publication, read and alignment data are deposited in sequence archives to enable other groups to reanalyze the data. While improvements in sequencing throughput and experimental protocols continue to generate ever-larger volumes of HTS data, pressing questions remain. Namely, how to store these data to minimize storage costs, maximize computational efficiency for data analysis, increase network transfer speeds to facilitate collaborative studies, or to facilitate reanalysis or perusal of data stored in archives.

A popular method for storing read data is in FASTQ files[5]. Such files are text files typically compressed with GZIP or BZip2 compression and hold both nucleotide bases of the reads as well as quality scores. The latter indicate the reliability of each base call and are central to many analyses, such as genotyping. Compressed text files have critical problems: they are slow to parse, and they do not support random access to subsets of the reads, a feature that is critical to support parallelization of the alignment of the reads to a reference genome.

Most analyses require aligning read data to a reference genome, a process that yields HTS alignment data. When computed, HTS alignment data has traditionally been stored in a variety of file formats. Early text formats were quickly abandoned in favor of binary formats, and among these, the BAM format has become very popular and is now widely used[6]. A key problem with BAM is that the BAM format cannot be seamlessly adapted to support new applications. For instance, developers of TopHat fusion were not able to extend the BAM format to store information about gene fusions, and instead had to create a variant of the SAM text format[7]. Developers of new analysis software based on BAM cannot seamlessly extend BAM for new applications because various programs that read/write BAM, developed worldwide, would need to be manually modified for each change to the specification, a process that is all but practical. Another key weakness of the BAM format is that BAM files require approximately the same amount of storage as the unaligned reads, for each alignment represented in the format.

The CRAM format, developed for the European Nucleotide Archive, was developed to try and compress HTS alignments better than can be achieved with BAM. CRAM achieves strong compression of alignment files using custom developed compression approaches parameterized on characteristics of simulated HTS alignment data[8]. A key innovation of CRAM was the recognition that different applications need to preserve different subsets of the data contained in a BAM file. Preserving different subsets of the information can yield substantial storage savings for these applications that do not require all the data. However, CRAM shares a key weakness with BAM. Namely, it is unable to seamlessly support changes to the data format. Changes to extend the file format require manual coding and careful design of custom compression approaches for the new data to be stored. An additional problem is that CRAM cannot compress HTS alignments when they are not already sorted by genomic position. This is a significant problem because alignments are first determined in read order before they can they be sorted by genomic position. This problem limits the usefulness of the CRAM format to HTS archives, and prevents its use as a full replacement of the BAM format. We believe that these weaknesses are serious drawbacks because the HTS field is progressing very rapidly, sequencing throughput increases exponentially and new experimental advances often require extensions to the data schemas used to store and analyze the new types of data.

In summary, the previous approaches are unable to strongly compress HTS data while supporting the full life-cycle of the data, from storage of sequenced reads to parallel processing of the reads and alignments during data analysis to archiving of study results. In this manuscript, we present a comprehensive approach that addresses these challenges simultaneously.

**Results**

**Structured data schemas.** We developed structured data schemas to represent HTS read and alignment data (Fig. S1A) with Protocol Buffers technology (PB)[9]. PB automates reading and writing structured data and provides flexibility with respect to changes in the schemas. Extending the schema requires editing a text file and recompiling the software. The new software is automatically compatible with versions of the software that are unaware of the schema extension. PB schema flexibility therefore provides the means to evolve the file formats over time as a collaborative effort without breaking existing software.

**Large datasets.** We extended PB with methods to store large datasets and to define configurable compression/decompression methods (called codecs, see Supp Methods). We developed codecs for general compression methods (PB data compressed with the GZip or BZip2 methods, Fig 1A), and a Hybrid codec that provides very strong compression of core alignment data, while retaining the flexibility of PB schema evolution (see Fig S1B and Supp Methods). Finally, we group data in tiers according to the most likely use of each kind of data (see Fig. 2) and have developed a framework and a set of tools (see [10] and Supplementary Materials and Methods) to support efficient computation with HTS data expressed in these formats. In the next sections, we describe our contributions to the compression of HTS structured data.

**General compression.** General compression approaches are widely used and were developed to compress unstructured data (e.g., streams of bytes such as text in a natural language). The most successful general compression approaches employ probabilistic compression, where smaller sequences of bits are used to represent symbols with high probability in the input data, and longer sequences are used to represent symbols of lower probability (the mapping from sequences of symbols to sequences of bits is called a code). Arithmetic coding can yield near-optimal codes (i.e., streams of bits of length close to the theoretical lower bound), given a model of symbol probabilities. However, the question of how to construct effective probabilistic models of unstructured data is a difficult one because the models have to be inferred from observing the stream of unstructured data. Since the cost of inferring the model grows with model complexity, progress in compression ratios has often been obtained at the expense of compression speed.

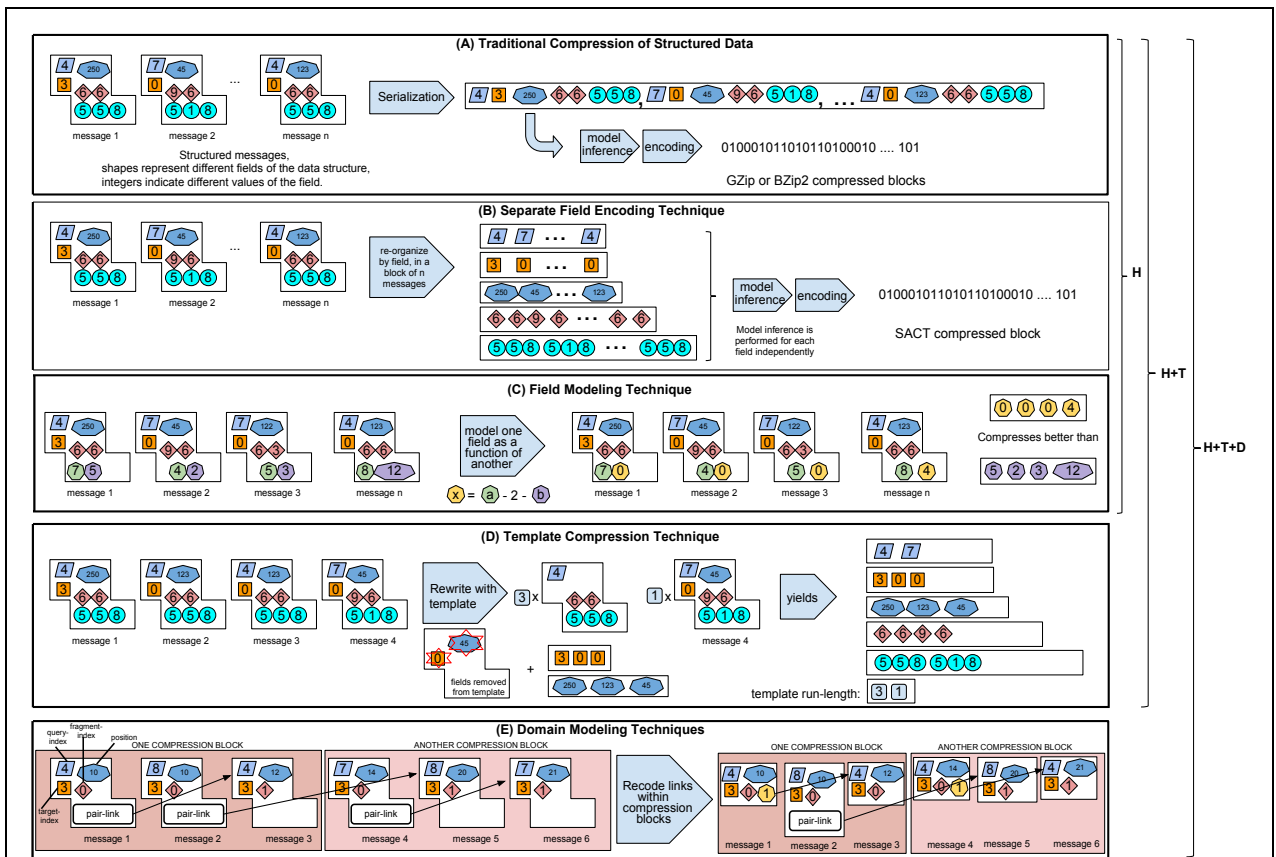

**Fig 1. Structured Data Compression Techniques.** We present the techniques that we devised for compressing structured High-Throughput Sequencing (HTS) data. We use a combination of general compression techniques (panel A) and of techniques that take advantage of the information provided by a data schema (B-E). (A) General compression techniques convert structured data to streams of bytes (serialization, typically done one message at a time) and then compressing the resulting stream of bytes with a general purpose compression approach such as Gzip and Bzip2. We use such techniques alone (Gzip and Bzip2 codecs) or in combination with structured data compression (Hybrid codecs, labeled H, H+T or H+T+D according to the technique used). (B) Separate field encoding reorganizes blocks of messages in lists of field values before compressing each field independently. The technique requires compressing blocks of PB messages, or PB chunks. (C) Field Modeling helps compress data by expressing the value of one field as a function of other fields and constants. (D) Template Compression Technique. Here, the data structure is used to detect subset of messages that repeat in the input messages. Fields that vary frequently are ignored from the template. The template values are stored with the number of template repetitions and the values needed to reconstruct the input messages. (E) Domain Modeling Technique. Alignment messages refer to each other with message links (i.e., references between messages) represented here as pair-link messages with three fields: position, target-index and fragment-index of the linked message. We realized that within a PB chunk, it is possible to remove the three fields representing the link and replace them with an integer index that counts how many messages up or down stream is the linked message in the chunk. Links from an entry in a chunk to an entry in another chunk cannot be removed and are stored explicitly with the three original fields.

**Structured data compression.** In this manuscript, we demonstrate that when presented with structured data (data organized according to a well-specified schema, see Fig S1(A) for an example), it becomes possible to leverage the data structure to facilitate model inference. We have devised and present several such techniques: separate field encoding, field modeling, template compression, and domain modeling (see illustration of these new methods on Fig. 1B-E).

**Separate Field Encoding.** This encoding reorganizes the dataset into lists of field values (Fig. 1B). Where a traditional approach to compressing structured data often applies a general compression method to a serialized stream of structured data (Figure 1(A)), we reasoned that compression could benefit from inferring a model for each field of a data structure separately (Figure 1(B)). Leveraging the structure of the data makes it possible for model inference to detect regularities in successive values of the same field. Field encoding requires compressing blocks of messages together. We call each such block a chunk of PB data and typically encode 10,000-100,000 messages per block.

**Field modeling.** This technique is useful when the value of one field can be calculated or approximated from the value of another field of the same data structure (Fig. 1C). In this case, this approach stores the difference between the approximated value and the actual value.

**Template compression.** This technique detects that a subset of a data structure (the template) repeats in the input (Fig 1D). It then stores the template, the number of times the template repeats and separately the fields that differ for each repetition. Template compression can be thought of as a generalization of run-length encoding to structured data.

**Domain modeling.** This technique requires a human expert to conduct a detailed analysis of the input data. In the case of HTS data, we developed an efficient representation of references between messages stored in the same PB chunk (Fig 1E).

**Evaluation.** We sought to evaluate the effectiveness of these approaches to compress HTS data by comparing them with general compression approaches (Gzip and BZip2) and with the BAM and CRAM approaches. To this end, because the compression effectiveness of any compression approach is expected to vary with input data, we assembled a benchmark of ten different datasets spanning several popular types of HTS assays: RNA-Seq (single-end and paired-end reads), Exome sequencing, whole genome sequencing (WGS), two DNA methylation assays: RRBS (reduced representation bisulfite sequencing) and whole genome Methyl-Seq (see Benchmark datasets in Materials and methods and Supp Table 1). Briefly, we found that combining all these approaches into the method labeled H+T+D (see Figure 1, H: Hybrid approach, T: template compression, D: domain modeling) yields the most competitive compression for HTS alignment data. Supplementary results provide details about how each technique affects performance (for instance, see general compression benchmark for comparison of the H and H+T methods with GZip and BZip2 compression in Table S2).

**General Compression Benchmark.** We evaluated the H and H+T approaches against general compression approaches when storing HTS alignment data. Table S2 presents the results of this benchmark. We tested the H and H+T variants of our approach (i.e., Fig 1B-D) and compared the evaluation metrics with those obtained when compressing the same data with the strong BZip2 general compression method [11, 12]. We measured file sizes and compression effectiveness, as well as compression and decompression times (Table S2) of the approaches compared. Each dataset is publicly available and all software required to replicate these results are freely available (see http://data.campagnelab.org/home/compression-of-structured-high-throughput-sequencing-data). Our benchmark results clearly establish that the H+T approach results in the smallest file sizes of either GZip or BZip2 compression. Compression ratios improve substantially when compared to BZip2 while compression/decompression speeds remain comparable or faster than BZip2. These results are significant because BZip2 is often considered as one of the most effective compression algorithms that remains sufficiently fast for practical use. Approaches that can achieve more than 10-15% better compression than BZip2 (i.e., 90-85% of the bzip2 baseline size, substantially larger than the results we report) are orders of magnitude slower for compression or decompression and are for this reason not practically useful for most applications. Using the H+T approach (Hybrid codec with Template compression, a lossless approach, see Methods), we show that we can compress structured alignment data to an average 57% of the size of BZip2 compressed data with similar computational cost.

**CRAM benchmark.** Table 1 compares performance metrics for storing alignments with the Hybrid codec or with CRAM. We find that the method H+T compression yields files comparable in size to the CRAM approach (average 101% of baseline). Interestingly, we noted that H+T compression is storing redundant information in the form of explicit forward and backward links between alignment entries (see pair links and splice links attributes in Fig S1A). The domain optimization technique described in Figure 1E, takes advantage of this observation and further increases compression of HTS alignment data (H+T+D approach) to an average 89% of the CRAM file sizes. Importantly, H+T+D appears to compress RNA-Seq and methylation HTS data much more effectively than CRAM (compressing these five datasets to 79% the size obtained with CRAM). Our results suggest that the CRAM approach may have been over-optimized to store WGS datasets to the extent that the optimizations become detrimental when compressing other types of HTS alignment data. Since the H+T+D approach adapts to the structure of the data, it does not suffer from this problem and yields substantial improvements when storing gene expression and epigenetic HTS datasets.

**Compression Fidelity.** When developing a new compression approach, it is critical to verify compression/decompression fidelity. We verified compression/decompression fidelity for the H+T and H+T+D approaches across all the benchmark datasets by comparing decompressed data to input data. When tested in the same conditions, we found that CRAM 0.7 decompressed three out of the 10 benchmark datasets to a size larger than that of the input BAM file (a situation that should never be observed with a compression approach), and failed to recover the exact value of specific fields for three additional datasets, indicating issues with the implementation we tested (see Table S4). These results suggest that the version of CRAM that we evaluated is not a completely mature HTS compression approach.

**HTS data analysis solutions.** A compression approach such as H+T+D is only part of a solution to support efficient analysis of HTS data. A comprehensive approach must also provide support for storing reads, storing alignments during their analysis (including unsorted alignments), and most importantly, it must provide tools that support the data formats and make it possible to conduct data analysis with compressed data.

**Comprehensive solutions for HTS data storage and analysis.** Figure 2 presents the elements of a comprehensive approach to HTS data management. These elements include the Goby framework[10], and extensions to widely used HTS analysis programs such as the GSNAP[13] and BWA aligners[14], or the Integrative Genomics Viewer[15]. Goby provides state of the art compression approaches, a rich application-programming interface to read and write HTS datasets and a program toolbox (see [10]). This toolbox can support the life-cycle of a HTS data analysis project, from alignment, to sorting and indexing the alignments by genomic position, to viewing aligned data in its genomic context with the Integrative Genomics Viewer.

**Multi-tier data organization.** These tools organize data in a multi-tier file organization. All reads (mapped or unmapped) are stored in Tier-I files and only differences between the reads and the reference sequence in alignment files (Tier II and III). This data organization reduces size considerably for Tier II alignments (in our benchmark up to 3-4% of the original BAM size for spliced RNA-Seq datasets, see Table 2), while keeping all the information that is typically stored in the BAM format in between the reads and alignment files. Combined, Tiers I and II capture all information currently stored in BAM files, but requires only about 50% of the storage capacity needed by the now popular FASTQ/BAM storage scheme (Table 2). The benefit of a multi-tier organization increases with the number of alignments performed against the same set of reads (Table 2). See Supp Methods for details about multi-tier data organization and further discussion of its advantages.

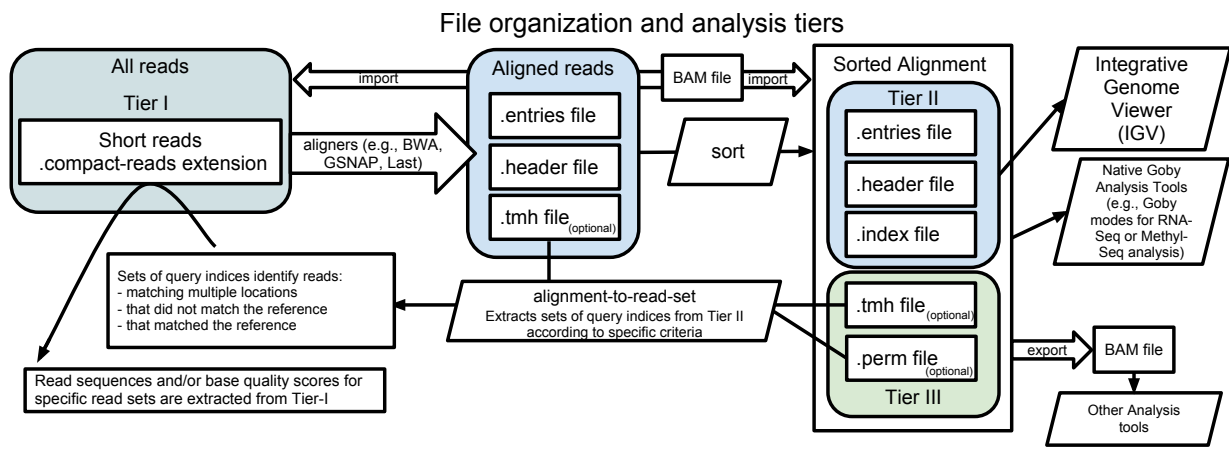

**Fig. 2. A comprehensive approach to store HTS data during the analysis life-cycle.** This diagram illustrates how HTS data stored with the approaches described in this manuscript support common analysis steps of a typical HTS study. HTS reads (Tier I) are stored in files ending with the .compact-reads extension. These files can be read directly by alignment programs and facilitate efficient parallelization on compute grids. When the reads are aligned to a reference genome, alignment files are written in sets of two or three files. Files ending with

.entries contain alignment entries. Each alignment entry describes how a segment of a read aligns against the reference genome. Files ending in .header contain global information about the reads, the reference genome, and the alignment (See Fig. S1(A) for the data schema that precisely describes these data structures). An optional .tmh file stores the identity of the reads that matched the reference so many times the aligner did not output matches for them. Aligned reads can be sorted with the 'sort' Goby tool, producing a .index file with enough information to support fast random access by genomic position. A permutation file (extension .perm) can also be produced to improve compression of sorted files (see Supp. Methods). Files in Tier II are stand-alone and can be transferred across the network for visualization (e.g., IGV). Files in Tier III are available for some specific types of analyses that require linking HTS alignments back to primary read data.

**Discussion**

We have presented new methods for storing, compressing and organizing HTS data. A key advantage of these methods is their ability to support seamless schema extensions, which makes it possible for data formats to evolve to meet the needs of the scientific community. These methods can be combined with compression methods to reduce the footprint of the datasets.

General compression methods such as GZip and BZip2 provide universal compression for any data schema. However, we have shown here that taking advantage of the structure of the data can yield state of the art compression of HTS alignment data. We anticipate that several of the techniques that we have introduced here can be generalized to arbitrary data schemas. For instance, separate field encoding, field modeling and template compression are techniques that could be used to develop fully automatic codec compilers (such compiler would analyze a data schema and yield a state of the art compression codec for the specific schema). We do not expect domain modeling to generalize to arbitrary schemas because it requires a detailed expertise about the data, but note that when used in combination with the other compression approaches described here, it has produced the new state of the art for compression of HTS alignment data.

The advantage of our approach may seem modest when compared to CRAM (11% smaller files on average over ten datasets). However, the compression advantage appears strongly dependent on the type of dataset compressed. We see a strong compression advantage for our approaches on RNA-Seq and RRBS datasets (average 21% advantage over 5 datasets). Similarly to CRAM, our approach requires the compression of blocks of HTS alignment data. CRAM uses blocks of one million entries. In our benchmark, we used blocks of 100,000 entries. We found that increasing block size improves the compression efficiency of our approach, however, large blocks also significantly slow down the performance of interactive visualization for the compressed alignments (e.g., visualization in IGV). To our knowledge, CRAM compressed alignments do not currently support interactive visualization, nor are any tools developed to directly analyze data represented in CRAM format. This is in contrast with the H+T+D approach that supports direct visualization and analysis with a number of HTS software tools[10]. Extending tools to support reading and writing the new file formats can be done by an experienced programmer in one or two days.

The multi-tier data organization that we introduce here makes it possible to reduce some of the computational burden of working with HTS data. Because Tier II alignment file sizes are greatly reduced, alignments can be transmitted through the network much faster than would be possible with the BAM format. This is a consequence of both better compression and multi-tier data organization. Multi-tier data organization facilitates visualization of the data as well as collaborative projects that require only access to aligned data (Tier II). For instance, alignments can be loaded through HTTP quickly with IGV. Multi-tier organization also improves the performance of cluster/cloud computing analysis pipelines that require staging of data on compute nodes before computations can take place.

Since Goby formats offer state of the art compression of HTS data and are seamlessly extensible for new applications, we propose that they are strong candidates to replace the FASTQ and BAM formats. We expect that the structured data compression approaches described here can be applied to a variety of scientific and engineering fields that need to store structured data.

**Acknowledgments**

Benchmarks datasets were constructed from public datasets (accession codes: NA12340, NA20766, NA18853, NA19172, GSM675439, GSM721194, ERP000765, ERP000765, SRR065390). We thank Dr. Olivier Elemento for critical feedback on revisions of this manuscript.

Methods are described in supplementary material.

**Table 1. Benchmark against a CRAM baseline**

| Sample ID | Kind | Compression ratio A/B= size(A)/size(B)*100% | |
|---|---|---|---|
| | | H+T/CRAM0.7 S2 | H+T+D/ CRAM 0.7 S2 |
| HZFWPTI | Exome | 101.76% | 87.92% |
| UANMNXR | Exome | 104.49% | 90.00% |
| MYHZZJH | RNA-Seq | 92.01% | 91.52% |
| ZHUUJKS | RNA-Seq | 90.73% | 88.70% |
| EJOYQAZ | RNA-Seq | 62.04% | 59.99% |
| JRODTYG | RRBS | 78.65% | 71.02% |
| ZVLRRJH | Methyl-Seq | 109.64% | 85.35% |
| XAAOBVT | WGS | 131.54% | 116.53% |
| UCCWRUX | WGS | 132.13% | 110.65% |
| HENGLIT | WGS | 109.22% | 92.07% |
| | Exome | 103.13% | 88.96% |
| | RNA-Seq | 81.59% | 80.07% |
| | bisulfite | 94.15% | 78.19% |
| | WGS | 124.30% | 106.42% |
| Average | All | 101.22% | 89.38% |

## Table 2. Benchmark against a BAM baseline

This table provides compression size ratios calculated for Tier 2 HTS alignments stored with approach H+T+D, with reads (Tier 1+2), or without (Tier 2 only). Reads are stored as PB data compressed with the BZip2 codec (R-BZIP2) or in FASTQ format compressed with Bzip2. The method H+T+D is configured to preserve soft-clips and their quality scores.

| Sample ID | Kind | Tier 2 Only: H+T+D / BAM | Tiers 1+2: (CR-BZip2 + H+T+D) / (FASTQ-Bzip2 + BAM) | Multi-tier storage with three alignments Tiers 1+2: (CR-BZip2 + 3 x H+T+D) / (FASTQ-BZip2 + 3x BAM) |
|---|---|---|---|---|
| HZFWPTI | Exome | 7.69% | 44.32% | 24.40% |
| UANMNXR | Exome | 7.45% | 44.17% | 24.19% |
| MYHZZJH | RNA-Seq | 2.83% | 29.14% | 13.31% |
| ZHUUJKS | RNA-Seq | 3.37% | 40.07% | 19.17% |
| EJOYQAZ | RNA-Seq | 13.00% | 54.54% | 32.97% |
| JRODTYG | RRBS | 24.21% | 69.82% | 46.28% |
| ZVLRRJH | Methyl-Seq | 26.99% | 61.75% | 43.12% |
| XAAOBVT | WGS | 8.15% | 49.67% | 27.87% |
| UCCWRUX | WGS | 6.85% | 49.93% | 27.59% |
| HENGLIT | WGS | 10.08% | 54.23% | 32.34% |
| **Average** | | **11.06%** | **49.76%** | **29.12%** |

# Compression of structured high-throughput sequencing data

Fabien Campagne[1,2]*, Kevin C. Dorff[1], Nyasha Chambwe[1,2], James T. Robinson[3], Jill P. Mesirov[3], Thomas D. Wu[4].

[1]The HRH Prince Alwaleed Bin Talal Bin Abdulaziz Alsaud Institute for Computational Biomedicine. [2]Department of Physiology and Biophysics, The Weill Cornell Medical College, New York, NY, USA. [3]Broad Institute of Massachusetts Institute of Technology and Harvard, Cambridge, MA, USA. [4]Department of Bioinformatics, Genentech, Inc., 1 DNA Way, South San Francisco, CA, USA.

*Correspondence to: Fabien Campagne, fac2003@campagnelab.org.

**Abstract**: Large biological datasets are being produced at a rapid pace and create substantial storage challenges, particularly in the domain of high-throughput sequencing (HTS). Most approaches currently used to store HTS data are either unable to quickly adapt to the requirements of new sequencing or analysis methods (they do not support schema evolution), or fail to provide state of the art compression of the datasets. We have devised new approaches to store HTS data that support seamless data schema evolution and compress datasets substantially better than existing approaches. Building on these new approaches, we discuss and demonstrate how a multi-tier data organization can dramatically reduce the storage, computational and network burden of collecting, analyzing, and archiving large sequencing datasets. For instance, we show that spliced RNA-Seq alignments can be stored in less than 4% the size of a BAM file with perfect data fidelity. Compared to the previous compression state of the art, these methods reduce dataset size more than 20% when storing gene expression and epigenetic datasets. The approaches have been integrated in a comprehensive suite of software tools (http://goby.campagnelab.org) that support common analyses for a range of high-throughput sequencing assays.

**Supplementary Materials**

- Tables S1-3
- Supp. Materials and Methods
- References
- Figures S1-S2

## Table S1. Details of the Benchmark Datasets

| Kind | BAM Tag | Paired-end | Spliced | Accession Code | Range | BAM file size | #reads | % mapped | Organism | URL |
|---|---|---|---|---|---|---|---|---|---|---|
| Exome | HZFWPTI | Yes | No | NA12340 | chr11 | 550 MB | 6,787,665 | 98.97% | human | [link] |
| Exome | UANMNXR | Yes | No | NA20766 | ch11 | 501 MB | 6,139,330 | 99.19% | human | [link] |
| RNA-Seq | MYHZZJH | No | Yes | NA18853 | all | 2.8 GB | 19,365,426 | 98.64% | human | |
| RNA-Seq | ZHUUJKS | No | Yes | NA19172 | all | 1.5 GB | 18,749,217 | 97.66% | human | |
| RNA-Seq | EJOYQAZ | Yes | Yes | | whole genome | 904 MB | 15,693,880 | 100.00% | human | [link] |
| RRBS | JRODTYG | No | No | GSM675439 | representative genome | 1.1 GB | 41,186,902 | 81.68% | mouse | [link] |
| Methyl-Seq | ZVLRRJH | No | No | GSM721194 | whole genome | 1.6 GB | 19,999,974 | 100.00% | human | [link] |
| WGS | XAAOBVT | Yes | No | ERP000765 | FC1, ~ first 20M reads | 1.4 GB | 19,999,953 | 100.00% | human | [link] |
| WGS | UCCWRUX | Yes | No | ERP000765 | FC2, ~ first 20M reads | 1.4 GB | 19,999,953 | 98.17% | human | [link] |
| WGS | HENGLIT | Yes | No | SRR065390 | First 30 million reads | 1.7 GB | 30,000,000 | 93.06% | c Elegans | [link] |

**Table S2. Benchmark against general compression baselines**
Compression size ratios A/B are calculated as the size of a dataset compressed with method A over the size of the same dataset compressed with method B, expressed as a percentage (a ratio A/B of 50% indicates that method A compressed the dataset to half the size of method B). Compression/Decompression speed ratios A/B measure the ratio of the time it takes method A to compress/decompress a dataset compared to the time it takes method B for the same dataset. A compression ratio A/B of 2.0 indicates that method A takes twice as long to compress a dataset than method B does. See Fig. 1 for a description of the H, H+T and H+T+D methods.

| Sample ID | Kind | Compression size ratios | | | Compression speed | | | Decompression speed | | |
|---|---|---|---|---|---|---|---|---|---|---|
| | | H / BZIP2 size * 100% | H+T / BZIP2 size * 100% | H+T / GZIP size * 100% | H/ BZIP2 | H+T/ BZIP2 | H+T/ GZIP | H/ BZIP2 | H+T/ BZIP2 | H+T/ GZIP |
| HZFWPTI | Exome | 66.53% | 57.57% | 40.06% | 0.84 | 0.96 | 2.29 | 1.54 | 1.49 | 3.47 |
| UANMNXR | Exome | 65.52% | 57.20% | 40.13% | 0.94 | 1.06 | 2.32 | 1.58 | 1.71 | 3.79 |
| MYHZZJH | RNA-Seq | 30.33% | 23.60% | 20.71% | 0.29 | 0.27 | 2.3 | 1.03 | 0.94 | 0.93 |
| ZHUUJKS | RNA-Seq | 33.32% | 26.43% | 22.97% | 0.44 | 0.4 | 1.99 | 1.49 | 1.31 | 2.73 |
| EJOYQAZ | RNA-Seq | 67.78% | 55.53% | 41.22% | 0.52 | 0.54 | 1.98 | 1.32 | 1.28 | 3.09 |
| JRODTYG | RRBS | 174.91% | 88.47% | 59.36% | 0.45 | 0.34 | 2.19 | 2.18 | 0.96 | 1.99 |
| ZVLRRJH | Methyl-Seq | 180.13% | 95.77% | 46.32% | 0.75 | 0.78 | 2.41 | 2.45 | 1.16 | 2.64 |
| XAAOBVT | WGS | 98.40% | 57.20% | 37.57% | 0.7 | 0.76 | 2.02 | 1.07 | 0.94 | 2.33 |
| UCCWRUX | WGS | 82.51% | 51.17% | 34.67% | 0.67 | 0.63 | 2.28 | 1.26 | 0.98 | 2.61 |
| HENGLIT | WGS | 66.21% | 59.01% | 41.53% | 0.91 | 1.29 | 3.19 | 1.44 | 1.27 | 3.79 |
| **Average all datasets** | | **87%** | **57%** | **38%** | **0.53** | **0.96** | **3.60** | **1.25** | **0.99** | **2.80** |

## Table S3. Compression/Decompression fidelity

This table presents round-trip compression/decompression ratios for each method evaluated. The ratio is calculated as the percentage of input file size restored after decompression. We compress the input BAM file with each method, then decompress the data back to BAM. Lossy decompression ratio must be below 100% since no decompressed file should be larger than the compressed file. Three datasets (numbers in bold) compressed with CRAM 0.7 S2 or S3 decompress larger than the input BAM file (see table S4). Most CRAM decompressed files add to the output a constant MQ BAM attribute that was not contained in the input files.

| tag | type | H+T+D | CRAM 0.7 S2 | H+T+Q | CRAM 0.7 S3 |
|---|---|---|---|---|---|
| HZFWPTI | Exome | 40.77% | 41.47% | 95.57% | 95.56% |
| UANMNXR | Exome | 40.33% | 40.97% | 95.67% | 95.65% |
| MYHZZJH | RNA-Seq | 27.98% | **71.62%** | 88.13% | **131.87%** |
| ZHUUJKS | RNA-Seq | 28.78% | **133.72%** | 88.65% | **194.49%** |
| EJOYQAZ | RNA-Seq | 27.60% | **1952.83%** | 84.96% | **2022.11%** |
| JRODTYG | RRBS | 48.77% | 47.61% | 73.39% | 71.32% |
| ZVLRRJH | Methyl-Seq | 60.26% | 59.56% | 87.54% | 86.08% |
| XAAOBVT | WGS | 45.28% | 45.95% | 90.15% | 90.42% |
| UCCWRUX | WGS | 39.67% | 40.62% | 88.86% | 89.54% |
| HENGLIT | WGS | 46.03% | 47.30% | 95.51% | 95.92% |

## Table S4. Compression/Decompression fidelity issues noted with CRAM

This table describes the discrepancies we observed when compressing BAM files with CRAM and decompressing them to BAM format.

(1) CRAM->BAM stores splices as deletions, which are decompressed with full explicit intronic sequences. Spliced BAM files decompress larger than source BAM file.
(2) Software failure when converting back to BAM
(3) No quality scores (all quality values restored as "?" in SAM file).
(4) It appears that CRAM calculates a value for TLEN (Insert Size) that does not always reproduce the TLEN value of the records in the input BAM file.
(5) CRAM outputs an MQ:I attribute when none was present in the input file.

| TAG | Type | CRAM 0.7 S1 | CRAM 0.7 S2 | CRAM 0.7 S3 |
|---|---|---|---|---|
| HZFWPTI | Exome | (3), (4) | (4) | (4) |
| UANMNXR | Exome | (3), (4) | (4) | (4) |
| MYHZZJH | RNA-Seq | (2), (5) | (1), (5) | (1), (5) |
| ZHUUJKS | RNA-Seq | (2), (5) | (1), (5) | (1), (5) |
| EJOYQAZ | RNA-Seq | (1), (3) | (1), (4), (5) | (1), (4), (5) |
| JRODTYG | RRBS | (3), (5) | (5) | (5) |
| ZVLRRJH | Methyl-Seq | (3), (5) | (5) | (5) |
| XAAOBVT | WGS | (3), (5) | (5) | (5) |
| UCCWRUX | WGS | (3), (5) | (5) | (5) |
| HENGLIT | WGS | (2), (5) | (4), (5) | (4), (5) |

# Materials and Methods

The HTS data management approach that we have developed combines novel methods (described in Figure 1 and main text of the manuscript) with a number of standard engineering techniques applied to HTS data:

- Schemas to organize HTS reads and alignments as structured data (Fig. S1A)

- The Protocol Buffers (PB) middleware to automate reading and writing structured data and to provide flexibility with respect to changes in the schemas [1].

- A storage protocol to store collections of billions of structured messages and support semi-random access to the messages in a collection. The protocol makes it possible to implement Codecs that compress/decompress the PB encoded collections. (Fig. S1B)

- GZip and BZip2 codecs

- A Hybrid codec that provides state of the art compression of alignment data, while retaining the flexibility of PB schema evolution.

- A multi-tier data organization that groups data in tiers according to the most likely use of each kind of data (Figure 2).

- A framework (see http://goby.campagnelab.org/) to support efficient computation with data expressed in these formats.

- A set of tools to work with reads and alignment data in these formats, including

  - Tools to import/export alignments from/to the BAM format.
  - Tools to import alignments written in the MAF format (produced by the Last aligner [2]).
  - Tools to import/export reads from/to the FASTA/FASTQ/CSFASTA formats for single-end or paired-end data [3].
  - Integrations into the BWA [4] and the GSNAP [5] aligners to natively load short reads and produce alignment results in these formats.
  - Various tools to help process alignments, including sorting, indexing, concatenating several alignments, or merging alignments performed against different reference sequences [3].
  - Extensions to the Integrative Genome Viewer (IGV [6]) to load and display the alignment format and display these data along side other data sources in a genomic context.

**Protocol Buffers.** Protocol Buffers (PB), developed by others, is a software middleware designed "to encode structured data in an efficient yet extensible format" (see http://code.google.com/p/protobuf/). PB offers data representation capabilities similar to the Extensible Markup Language (XML), but that are simpler and significantly more computationally efficient. PB schemas provide a formal data representation language that can express primitive language types as well as complex types of data and their relationship to form messages (equivalent to objects in an OO language or structures in languages such as C). PB provides compilers for a variety of languages that transform schemas into program components suitable to represent data in memory, as well as serialize these data (i.e., write messages to a buffer of bytes) or de-serialize messages (i.e., read a buffer of bytes to reconstitute well-formed messages). Figure S1A presents the PB schemas we devised for storing reads and alignments, respectively. The latest schemas can be obtained at http://github.com/CampagneLaboratory/goby/blob/master/protobuf/Alignments.proto and http://github.com/CampagneLaboratory/goby/blob/master/protobuf/Reads.proto. We use Unified Modeling Language (UML) conventions similar to those described in [7] to document the relationships between messages used to store reads or alignments. Rather than encoding complex information in strings (e.g., sequence variations stored as "CIGAR" strings in BAM), we decompose the information into different PB messages that are simple to process computationally.

Protocol Buffers were initially developed to transmit small messages in client-server environments where software needs to be upgraded asynchronously. As such they provide strong capabilities for schema evolution. For instance, it is possible to add a new message field to a copy of the AlignmentEntry schema, write software that populates the new field, and send data files with this new schema to third parties. Such third parties will be able to use older versions of the software to extract all but the new data structure in the data files. Reading data from the new field will require new software, but third parties can decide if and when they upgrade.

Because PB was initially developed for small messages it would be natively unsuitable for serializing or deserializing the very large collections of messages needed to store billions of reads or alignments. We introduce a storage protocol that addresses this deficiency while retaining the schema evolution capabilities of PB.

**Large Collection Storage Protocol.** To work around the message size limitation of PB, we introduce the Goby Large Collection Storage Protocol (GLCSP), depicted in Fig. S1B. Briefly, this protocol represents collections of PB messages with N elements as K collections of N/K messages. In the benchmark, we used K=100,000 so large collections are represented with chunks that contain at most 100,000 messages. Large collections are represented as sequences of chunks. GLCSP supports semi-random access since it is possible to start reading into a GLCSP formatted file at any position and scan until the start of another chunk is encountered. The next chunk found can be decompressed with the codec associated with the registration code found at the start of the chunk (an error to decompress the chunk indicates a false positive delimiter detection, which will be statistically quite infrequent, but needs to be handled appropriately to resume scanning for the next chunk). A termination chunk with 8 successive 0xFF bytes followed by four 0x00 bytes is written immediately before the end of file.

**GZip codec.** The GZip codec, used since Goby 1.0, simply encodes PB serialized data with GZip compression. The implementation of the GZip codec in Goby uses the standard Java classes: java.util.zip.GZIPInputStream and java.util.zip.GZIPOutputStream.

**BZip2 codec.** The BZip2 codec, introduced in Goby 2, encodes PB serialized data with BZip2 compression. The implementation of the BZip2 codec in Goby uses the Apache ant implementation: classes named org.apache.tools.BZip2.CBZip2InputStream and org.apache.tools.BZip2.CBZip2OutputStream.

**Hybrid Codec.** The Goby 2 hybrid codec encodes PB serialized data in two serialized streams: the ACT stream and the Left-Over PB stream. ACT stands for Arithmetic Coding and Template. This new compression approach, described below, consists of compressing collections of K structured messages by serializing fields of the structure messages with arithmetic coding compression. An ACT codec must be implemented for each different kind of schema. At the time of writing this report, we have implemented an ACT codec for the alignment schema described in Fig. S1A. Since PB supports seamless evolution of PB schemas, the hybrid codec must be able to store data that could not be handled by a given ACT implementation (for instance data from fields that have been added to the schema after the ACT implementation was compiled, possibly by a third-party). The hybrid codec stores such data in the Left-Over PB stream. To this end, PB data is serialized and compressed with GZip (Fig. 1A).

**Arithmetic Coding and Template (ACT) Compression.** This approach takes as input a collection of structured messages and produces a stream of bytes with compressed data. This is achieved by considering each field of the messages independently and collecting the successive values of the field when traversing the collection from the first PB message to the last. We reduce each field type to a list of integer. Such lists are compressed as described in section Integer List encoding. Field types are handled as follows. Fields that are recognized by an implementation of ACT (produced against a specific version of the data schema) are removed form the input PB message. Input PB messages that remain non-empty after processing all fields that the codec is aware are written to the Left-Over PB collection output. This simple mechanism suffices to guarantee that older versions of the software do not erase new data fields needed by more recent versions of the software. Fields that are recognized by an ACT implementation are processed as follows, according to their type:

*Integer fields.* Fields that have a small number of distinct values across all elements of the input collection are written with arithmetic coding list compression (see below). We first introduce the coding techniques used by our integer list compression approach. Fields that follow a uniform distribution (i.e., queryIndex) are written with minimal binary coding.
*String fields.* String values are converted to list of integers by successively encoding the first character of each string field, then the second, and so on until the length of each string is reached. The length of each string is recorded in a separate integer list.
*Floating number fields.* Floating numbers (32bits) are stored as their integer representation in an integer list.

**Cost of Model Inference.** It is important to note that ACT does not eliminate the cost of model inference. Where other approaches incur this cost when presented with a new data file, ACT incurs most the cost once for every data schema, and a much smaller cost for each dataset (for instance when deciding to use run-length encoding for a field of a given dataset). The cost of model inference incurred for the schema is thus amortized over many data files represented with the schema.

**Run Length Encoding.** Integer lists are scanned to determine if run-length encoding would be beneficial. To this end, a 'lengths' and 'values' list is created from each integer list to code. The 'length' list stores the number of times a given value repeats. The 'values' list simply contains the values of the input list. When the sum of the 'lengths' and 'values' list is smaller than the input list, run-length encoding is used (i.e., we separately write 'lengths' and 'values' lists as described in the section Integer List Compression). Otherwise, the input list is written directly with as a list of integers (see Integer List Compression).

**Nibble Coding.** Nibble coding is a variable length encoding technique that represents small integers with a small number of bits. We use the Nibble coding implementation provided in the DSI utilities (http://dsiutils.dsi.unimi.it/, Sebastiano Vigna and Paolo Boldi). The following description is copied from the documentation of the DSI package. Nibble coding records a natural number by padding its binary representation to the left using zeroes, until its length is a multiple of three. Then, the resulting string is broken in blocks of 3 bits, and each block is prefixed with a bit, which is zero for all blocks except for the last one.

**Minimal Binary Coding.** A minimal binary code is an optimal code for the uniform distribution and is used to encode query indices (used to link alignment data to read data, see multi-tier organization sections). Briefly, knowing the range of values to be encoded, one can write a natural number in binary code using m bits, where m suffices to encode the maximum value. The value of m is determined by calculating the most significant bit of the maximum value of the list. Minimal binary coding is performed with the DSI utilities. Query indices are written as q-min_q, where q is a query index in a PB collection, and min_q is the minimum value observed in the same collection. The parameter m is detemined as max_q – min_q +1, where max_q is the maximum query index observed in the collection.

**Arithmetic Coding.** An arithmetic coder is a compression method that yields a code of near optimal length given a specific symbol probability distribution. Arithmetic coders can estimate symbol probabilities adaptively. We use an arithmetic coder implementation derived from that offered by MG4J [8]. However, the Goby implementations of the arithmetic decoder have been optimized for large symbol alphabets (the MG4J decoder has complexity of decoding a symbol $O(n)$, where n is the number of symbol, while the Goby implementation has complexity $O(\log(n))$.

**Integer List Compression with Arithmetic Coding.** Lists of integers are first inspected to determine if run-length encoding is beneficial. If this is the case, the list is processed as two lists as previously described. Each integer list is then encoded as follows. We write the number of elements of the list with nibble coding, followed by the sign bit (one zero bit if all symbol values are positive, or a one bit if they contain negative values), followed by the number of symbols (nibble coding), the value of each symbol (nibble coding after applying a bijection to map negative integers to natural numbers, when the sign bit was 1). The index of the symbol for each value of the list is then written in sequence using arithmetic coding.

**Boolean List Compression.** Booleans are converted to the integer value zero or one to produce a list of integers, and further processed as described in the previous section.

**Lists of Structured Messages.** PB supports messages that contain other structured messages. Goby schemas use this capability to encode sequence variations and links (see Supp. Fig. 1A). We compress lists that refer to other messages with as many lists of integers as required to compress each field of the linked message type. Additionally, for each source message, we store the number of elements of the destination message that belongs to the source. For instance, when an AlignmentEntry message includes three SequenceVariation messages, we add the number three to the list that stores the number of sequence variations per entry. We then inspect each SequenceVariation message in the order they appear in the sequenceVariations field and append to the list associated with each field (readIndex, position). Since the traversal order is fixed, this approach can reconstruct both links and linked objects by decoding three sequence variations from the field elements, consuming three readIndex or position values for the entry message.

**Template Compression.** Template compression is a generalization of the run-length encoding technique for input structured messages. Briefly, for each type of message, we choose the set of fields that will not be included in the template (*non-template fields*). These fields should be chosen as those fields that change the most from one entry to the next. The value of each non-template field is recorded to its respective integer field list and the field is removed from the PB message. In the current implementation, we remove the fields queryIndex, position and toQuality to yield the template. After removing all non-template fields, we are left with a template message. We check if the previously encoded message has the same value as the current template message and if yes increment the number of times the template is to be emitted (we do not emit individual fields for the template in this case). If not, we emit individual fields. A more formal description is given under section 'Algorithm template compression'.

**Benchmark datasets.** Benchmark datasets were obtained from public databases whenever possible. Accession codes are provided in Table S1. The larger files were trimmed to keep only about twenty million reads. Reads that did not map where filtered out. The exact reduced datasets used for the benchmarks can be obtained from http://data.campagnelab.org/

**Alignments.** Alignments were obtained in BAM format. For samples MYHZZJH and ZHUUJKS, we obtained reads from [9] and realigned against the 1000 genome reference sequence (corresponding to hg19) with GSNAP (version 2012.01.11), allowing for spliced alignments (options for de-novo and cDNA splice detection were enabled). The resulting BAM output is available from http://data.campagnelab.org/.

**Benchmark methodology.** We developed a set of Bash and Groovy scripts to automate the benchmarks. These scripts are distributed with the benchmark datasets to make it possible to reproduce our results and to assist with the testing and development of new codecs. Scripts copy all data files to local disk before timing execution, and write results to local disks as well, to remove possible variability induced by network traffic. Compression ratio A/B for methods A and B are calculated as the file size obtained when compressing a benchmark file with method A divided by the file size obtained when compressing the same file with method B. Compression ratio are shown as percentage, where 50% indicates that method A compresses the data to 50% the size achieved by the baseline method B. Compression ratios are deterministic, so we do not repeat these measures. Compression speed is measured with the Linux *time* command (using 'real time'), subtracted with the time taken to read the same input file on the same machine (this time is measured as the time taken to compress the input with a 'null' codec, a codec that writes no output). Decompression speed is measured as the execution time of the Goby compact-file-stats mode. This mode decompresses the successive chunks of an alignment file and estimates and reports statistics about the entries in the alignment. Compression and decompression speeds are largely deterministic, varying only by a few percentage from run to run. We omitted standard errors for clarity because repetitive runs showed virtually no variation in our test environment.

**CRAM parameter settings.** We contacted the authors of the CRAM compression toolkit to request that they suggest three parameter settings to evaluate their tool in a comparative benchmark. They suggested setting S1: the most lossy compression, does not keep soft clips nor unmapped placed reads (command line arguments: `--ignore-soft-clips --exclude-unmapped-placed-reads`). Setting S2: intermediate lossy compression, keeps soft clips and unmapped placed reads, keeps quality scores for mutations and insertion deletions (command line arguments: `--capture-substitution-quality-scores --capture-insertion-quality-scores`). Setting S3: lossless compression, like S2 but also keeps quality scores for the complete reads, the BAM attribute tags and preserve unmapped reads (command line arguments: `--capture-all-tags --capture-all-quality-scores --include-unmapped-reads`).

**Goby parameter settings.** BAM files were converted to Goby file format with the sam-to-compact tool. Files were initially written with the GZIP codec, and re-compressed with each codec using the concatenate-alignment tool (see benchmark scripts). We used the following parameters to measure compression with the ACT approach:

To create files comparable with CRAM0.7 S2, we preserved soft-clips and quality scores over variations (option `--preserve-soft-clips` of the Goby sam-to-compact tool). To create files comparable with CRAM0.7 S3, we preserved soft-clips and quality scores over the entire read (options `--preserve-soft-clips --preserve-all-mapped-qualities` of the Goby sam-to-compact tool).

The following settings were used with the concatenate-alignment tool:

- Compression with gzip codec: `-x MessageChunksWriter:codec=gzip -x AlignmentWriterImpl:permutate-query-indices=false -x AlignmentCollectionHandler:ignore-read-origin=true --preserve-soft-clips`.
- Compression with bzip2 codec: `-x MessageChunksWriter:codec=bzip2 -x AlignmentWriterImpl:permutate-query-indices=false -x AlignmentCollectionHandler:ignore-read-origin=true`
- Compression with ACT H approach: `-x MessageChunksWriter:codec=hybrid-1 –x MessageChunksWriter:template-compression=false -x AlignmentCollectionHandler:enable-domain-optimizations=true –x AlignmentWriterImpl:permutate-query-indices=false -x AlignmentCollectionHandler:ignore-read-origin=true`
- Compression with ACT H+T+D approach: `-x MessageChunksWriter:codec=hybrid-1 –x MessageChunksWriter:template-compression=true –x AlignmentCollectionHandler:enable-domain-optimizations=true -x AlignmentWriterImpl:permutate-query-indices=false -x AlignmentCollectionHandler:ignore-read-origin=true`

**Multi-tier data organization.** A critical advantage of a multi-tier file organization is the ability to study a dataset with multiple alignment methods. With single file organization (e.g., BAM or lossless CRAM), multiple analyses result in duplicating data. Indeed, projects that produce BAM alignments typically already have read data in FASTQ format. The read data are duplicated in each new BAM file that a project produces with a different alignment approach. With multi-tier organization, only alignments are stored for each analysis, further increasing storage efficiency for projects that align reads with multiple methods. Table 2 indicates that multi-tier organization can yield substantial storage savings when compared to a FASTQ/BAM storage scheme. It is worth noting that each schema includes fields that provide meta-data to assist tracking relations between data files (See Fig S1A, MetaData message in the read schema, and ReadOriginInfo in the alignment schema).

**Preserving read trackability.** Multi-tier organization must preserve the identity of a read across tiers. This is necessary to link back specific analysis results to raw data. In the BAM format, read-names link alignment results to the primary read data. This solution is effective, but wasteful: it requires storing long strings of characters (~15 characters in most current datasets, or 6+ bytes at least if unique integers are written as strings) whose only function is to maintain read identity. In the Goby multi-tier organization, read identity is maintained with an integer index. This index tracks read identity during the entire life cycle of HTS data. Special considerations must be taken to guarantee that preserving this index during the data life-cycle does not degrade compression performance. We discuss these methods in the next section.

**Query Index and Permutations.** Goby maintains the link between alignments and primary read data with an integer, called a query index (See Fig. S1(A), ReadEntry message type). When an alignment program processes a Goby reads file, the query index field of the read entry is written to the alignment entry to preserve the link to the raw data (see Fig. 1 (B), AlignmentEntry). Sorting the alignment will result in shuffling query indices, and can seriously degrade compression performance of a sorted alignment (because compression of a sequence of uniformly distributed 32 bit integers requires 32 bits per integer). We avoid this problem by permuting the original query indices to small indices that monotonically increase in genomic order. The small indices are still uniformly distributed, but in a much reduced range, and therefore can be compressed more effectively (with the minimal binary coding method). Permutations are written to disk in a specialized data structure that makes it possible to retrieve the original query index corresponding to any small index (stored in a .perm file in Tier III). Permutation files are only necessary for those applications that need to track read indices back to primary read data. Tier II alignments can be used in isolation or together with data from Tier III, depending on the needs of the application. We note that CRAM did not address the issue of storing read identity because (1) CRAM can only compress sorted alignments, (2) Converting a BAM file to CRAM does not maintain a mapping between read index and read name.

**Compression/decompression fidelity.** We tested whether each compression approach was able to recover the input dataset after decompression. A simple test of compression/decompression fidelity is to compare the size of the dataset decompressed after compression with an approach. We found that CRAM yielded datasets larger than the input BAM file for three of the ten benchmark datasets (See Supp Table 2). In one instance (dataset EJOYQAZ), the extracted BAM file was 20 times larger than the input BAM file because of incorrect output of spliced alignments. A more stringent test consists in comparing input and decompressed data files byte for byte. This approach is not practical when the compression approach preserves only a subset of the input file. To work around this problem, we visually inspected subsets of the alignment file in IGV. We found that CRAM0.7 S1 (the most lossy parameters for the CRAM approach, see CRAM parameter settings section) introduced many spurious sequence variations throughout the alignment in some reads. Importantly, S1 also did not appear able to preserve quality scores for sequence variations, which makes the setting of questionable practical use. Because of these serious fidelity problems and limitations, we do not report results for CRAM 0.7 S1 setting. We found a few additional fidelity problems with CRAM 0.7 S2 or S3, described in Table S4. We verified that the approaches H+T or H+T+D had no fidelity issues on the same datasets.

**<u>Algorithm Coding and Template compression:</u>**

```
Inputs:
     A collection of PB messages in a GLCSP chunk
Outputs:
     A number of integer lists, one for each field of the PB
     messages
Init:
T=nil
C=0
for each message m in PB chunk:
     for each non template field g of m:
          emit g to integer list corresponding to g field
          remove g from m
     end
     if g equals T then
          C+=1
          continue with next message
     else
          emit C to integer list corresponding to message count
field
          for each field e of m:
               emit e to integer list corresponding to e field
          end
          T=m
          C=1
     end
end
```

# Supplementary Figures

Fig. S1. Structured data schemas and Large Collection Storage Protocol.
Fig. S2. Fidelity issues with CRAM 0.7 S1 compression/decompression.

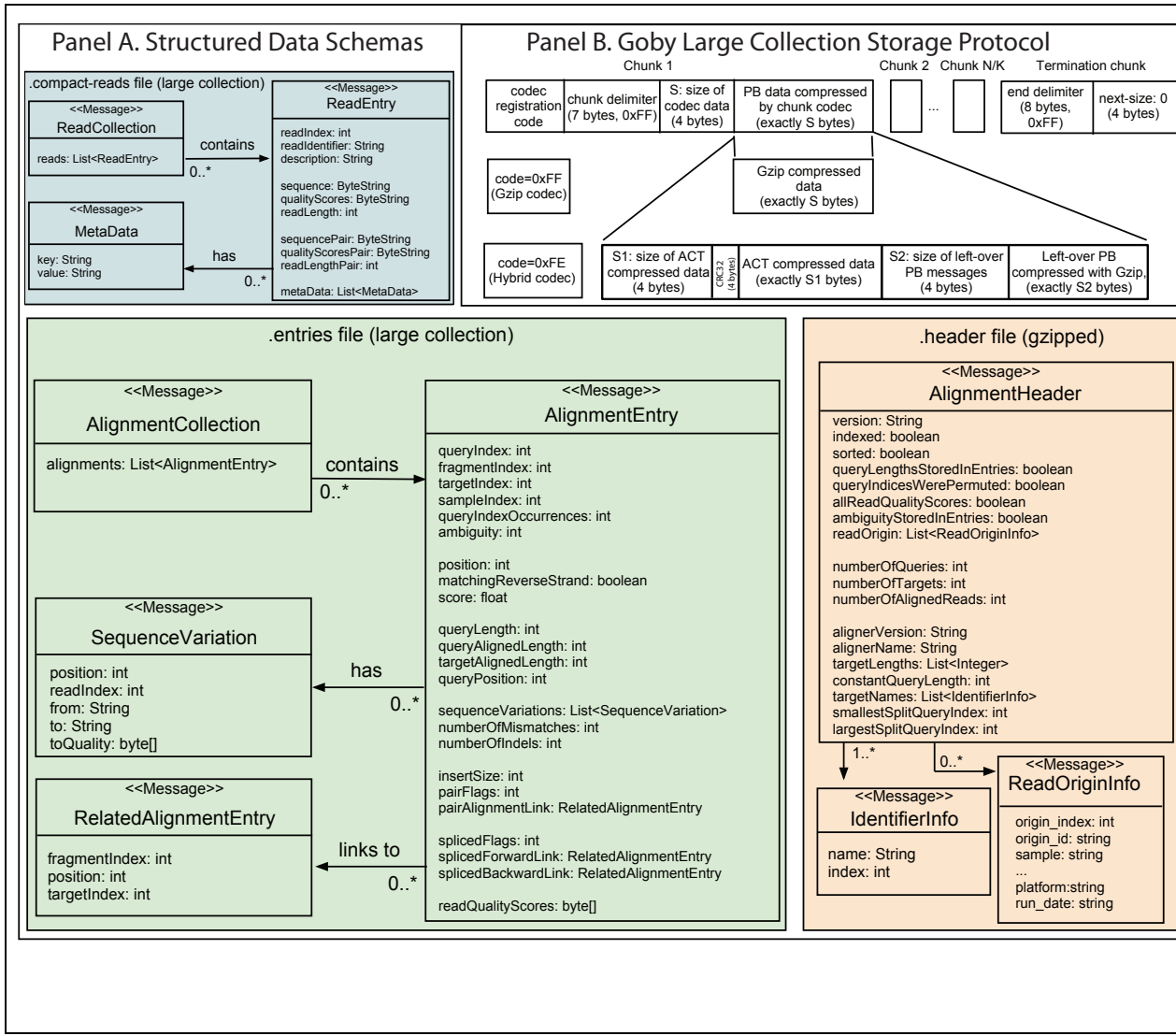

**Figure S1. Structured data schemas and Large Collection Storage Protocol.** (**A**) Describes the data schemas that we have devised to represent HTS reads and alignments. These schemas are described following the Unified Modeling Language conventions. Briefly, data are organized as Protocol Buffer messages (equivalent of data structures), which can contain data fields of primitive types or other messages. We store reads as collections of ReadEntry messages and alignments as collections of AlignmentEntry messages. References between messages are represented with integer indices. For instance, the RelatedAlignmentEntry message is used to link two alignment entries to represent paired or spliced alignments. (**B**) Describes how large collections of messages are stored piecewise in chunks of compressed PB data. The Goby Large Collection Storage Protocol (GLCSP) provides a plugin mechanism to define new PB compression/decompression approaches (codecs). The Hybrid codec compresses one part of a collection with the ACT approach (H, H+T or H+T+D variants) and the reduced messages remaining after ACT compression with GZip.

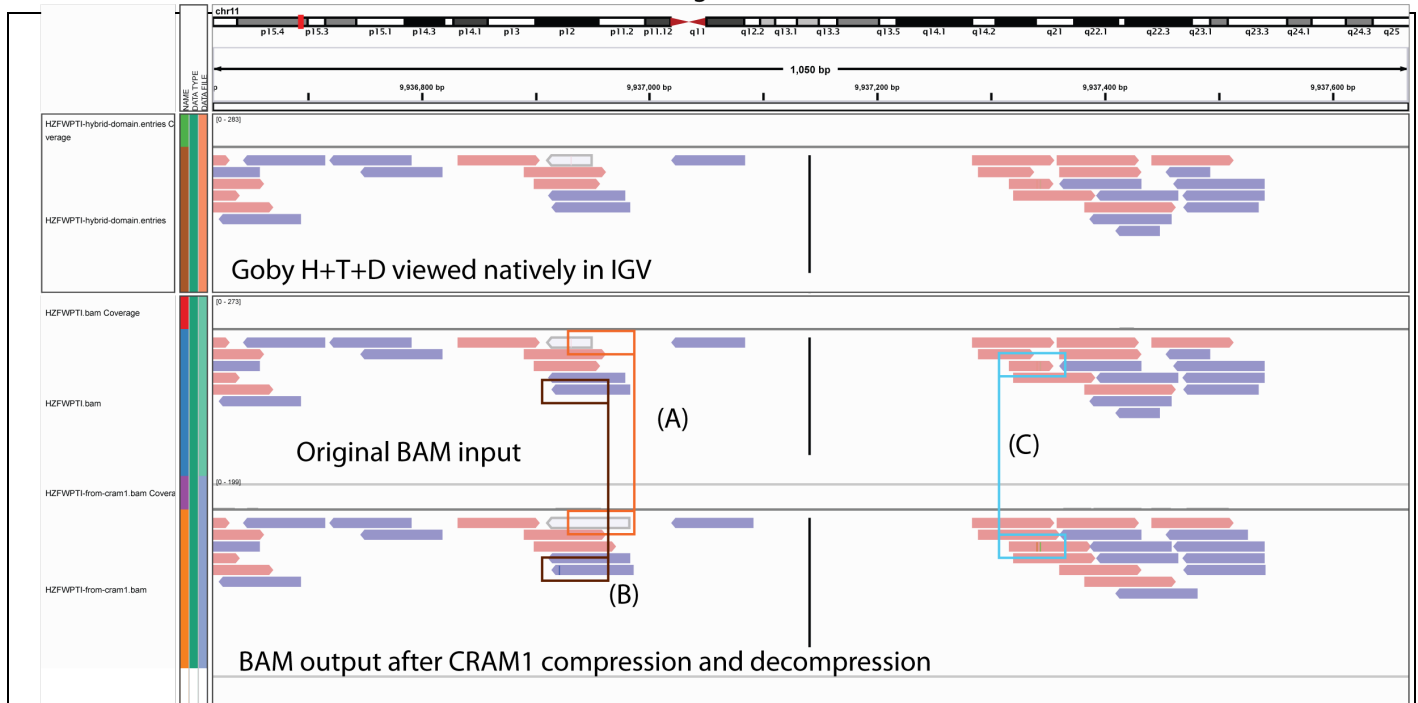

**Figure S2. Fidelity issues with CRAM 0.7 S1 compression/decompression.** When comparing an input BAM file to the BAM file generated after a round-trip compression/decompression with CRAM 0.7 S1, we noted three kinds of issues. **(A)** Some alignment block boundaries are inappropriately conserved, in the region highlighted, the CRAM1 BAM output extends the end of the alignment block. Futhermore, in this block, one sequence variation is not reproduced, while another variation if introduced in the middle of the block that did not exist in the input. **(B)** Another example of sequence variation being introduced that did not exist in the input. **(C)** While the sequence variations are preserved (i.e., position and type of base change), the CRAM decompressed BAM file lacks all quality scores for these variations. We conclude that the S1 setting of CRAM 0.7 has too many issues to be practically useful. This example can be examined in more detail in IGV using the supplementary online information available from http://data.campagnelab.org.